\documentstyle[aps,prl,eqsecnum]{revtex}

\begin{document}
\title{Weak magnetohydrodynamic turbulence of magnetized plasma}
\author{E.A.Kuznetsov\footnote{ e-mail: kuznetso@itp.ac.ru}}
\address{{\it Landau Institute for Theoretical Physics,
 2 Kosygin str., 117334 Moscow, Russia}}

\maketitle

\begin{abstract}
Weak turbulence of magnetohydrodynamic (MHD) waves in strongly 
magnetized plasma 
is studied when the plasma pressure is less than the magnetic field pressure.
In this situation the main nonlinear mechanism is the resonance scattering
of fast magneto-acoustic and Alfvenic waves on slow magneto-acoustic waves.
As a result, the former waves serve as the high-frequency waves with respect to
the latter ones so that the total number of HF waves - an adiabatic invariant - 
conserves additionally.  
In the weak turbulence regime this invariant is shown to 
generate the Kolmogorov type spectrum with a constant flux of
HF waves towards large-scale region. In the short-wave region 
another Kolmogorov spectrum can be realized with a constant energy flux.
The explicit angle dependences  for both types of turbulent spectra are 
found for 
the propagation angles close to the direction of a mean magnetic field.

\end{abstract}

\section{Introduction}

The central place in theory of turbulence 
occupies the notion of spectrum of turbulence -- the energy distribution 
on scales. The problem of its finding is one of the most difficult and nowdays
is far from the complete solution. For developed hydrodynamic turbulence 
the pioneering works of A.N.Kolmogorov \cite{kol} and A.M.Obukhov 
\cite{obukh} of 1941
about self-similar nature  of turbulent spectra determined 
development of theory of turbulence for many 
years. In the seventieth years mainly by the efforts of V.E.Zakharov
the Kolmogorov-Obukhov ideas were fruitfully applied to the weak wave 
turbulence
theory (for details, see monograph \cite{ZLF} and also the first papers 
devoted to this subject \cite{Z1,Z2,ZF}).  
Wave turbulence in some sense occurred to be simpler than developed hydrodynamic 
turbulence.
The main reason of such simplicity is connected with the wave dispersion when 
there exists such a region of  wave intensities  when the nonlinear 
interaction between waves can
be considered weak in comparison with dispersive effects. If initially 
phases of waves are distributed randomly 
then the nonlinear
interaction can provide a weak correlation in phases of the interacting waves. 
By this reason
such ensemble of waves can be described in terms of pare correlation functions 
the Fourier spectra of which coincide, up to the multiplier, with a number 
of waves
$n_k$ (occupation number) with the definite wave vector ${\bf k}$. 
In their turn,
the occupation numbers $n_k$ obey the kinetic equations for waves.
The Kolmogorov spectra in this theory arise as stationary scale-invariant 
solutions
of the kinetic equations annulating their collision terms. 
These spectra, unlike the thermodynamic equilibrium ones, can be related 
to the solutions with
nonzero fluxes over scales. They realize a constant flux over scales of some integral 
of motion:
energy, number of waves, etc., in the so-called
inertial interval.  It is important that if for developed hydrodynamic turbulence
the notion of the inertial interval -- the region where influence of both pumping and dissipation 
can be neglected -- in fact represents the hypothesis of locality, for the weak turbulence theory
this property can be checked explicitly. 

It is necessary to note that the main mass of works devoted to the Kolmogorov
spectra of weak turbulence consider isotopic media
(the corresponding bibliography can be found in \cite{ZLF}). Influence 
of anisotropy,
for instance, magnetic field in plasma, has been studied in the less extent. 
The first such example of  Kolmogorov type spectra was considered by the author in 1972 
\cite{kuz} for weak turbulence of magnetized ion-acoustic waves
in plasma. For this case  the collision term of the kinetic equation
was shown to be invariant with respect to two independent scalings along 
and perpendicular 
the magnetic field that allowed, by means of generalization of the
Zakharov transformation, to construct the Kolmogorov spectra with 
the power dependences relative to longitudinal ($k_z$) and transverse 
($k_{\perp}$) components of the wave vector. Later, the ideas of these 
papers were used for finding Kolmogorov spectra for drift waves in plasma and 
for the Rossby waves (see, for instance, \cite{nazar,nazar1}).

The present paper is devoted to weak turbulence of magneto-hydrodynamic 
waves in 
strongly magnetized plasma when the thermal pressure of plasma $nT$ is 
small compared with
the magnetic field pressure $H^2/(8\pi)$ $(\beta = 8\pi nT/H^2\ll 1)$. 
In this case,
in comparison with \cite{Z1}, \cite{kuz},   turbulent spectra are defined from 
solution of three linked kinetic equations for  
 Alfvenic, fast and slow magneto-acoustic waves.

For $\beta \ll 1$ the main nonlinear mechanism of the MHD waves is
a scattering
of fast magneto-acoustic and Alfvenic waves on slow magneto-acoustic waves 
(Section 2).
In these processes, which can be considered as a partial case of decays of 
one wave to two another waves and the oppsite process of fusion,  Alfvenic and 
fast magneto-acoustic 
waves play a role of high-frequency waves against slow magneto-acoustic waves. 
In each act of scattering the frequency change of the former waves 
(we will call them A-waves)
is small enough,  due to smallness of the parameter $\beta$. By this reason this process is 
familiar to the 
Mandelstamm-Brillouin scattering - scattering of electromagnetic waves 
on  acoustic phonons.
According to separation in time scales - devision of all waves onto  HF and LF 
waves -  
the decay interaction conserves, besides the energy, the adiabatic invariant -- 
the total number of HF waves. 
However, this analogy with  the Mandelstamm-Brillouin scattering 
has not been exhausted by the said above. It turns out 
that the matrix element of this interaction has maximum for
the maximal value of longitudinal projection of momentum transfered by 
A-waves to slow magneto-acoustic waves. In particular, 
this result can be extracted from the
expression  for the growth rate of  decay instability for the monochromatic
Alfvenic wave obtained in 1962 by A.A.Galeev and V.N.Oraevskii \cite{GO}.
Remind that for the  Mandelstamm-Brillouin scattering the matrix element 
is also proportional to square root from the value of transferred momentum,
that provides  maximum for the back scattering of electromagnetic waves. 
Because of such behavior of the scattering amplitude of A-waves it is 
natural to assume 
that a stationary distribution of waves over angles
will be very anisotropic, concentrated in the $k$-space along the magnetic 
field direction.
Under such assumptions the kinetic equations possess two additional symmetries
-- invariance with respect to two independent stretching along and in 
transverse direction to
the mean magnetic field
that allow one  to use the transformations of the paper \cite{kuz}.
Due to two these  symmetries  of the kinetic equations it turns out that in 
the transparency region there are possible
two scale-invariant (against longitudinal and transverse wave vectors) 
Kolmogorov spectra
corresponding to a constant flux of energy  to the short-wave region - the direct 
cascade  and to a
constant flux of A-waves towards the large-scale region - the inverse cascade. 
This paper is based on the old results of the author  \cite{kuz1} published
in the form of preprint in Russian and unknown by this reason abroad (that 
is natural).
But it turns out that these results are also unknown for Russian readers.  In spite 
of more than twenty five
years history nobody has not repeated these results. It is necessary to note, 
however, 
that recently the question about MHD turbulence has been considered
in another limit $\beta \gg 1$ in \cite{nazar2}.
This limit differs significantly from that considered in the present paper. 
First, at  $\beta \gg 1$ plasma can be considered as almost incompressible 
fluid and,
secondly, in this limit there is no essential difference  between Alfvenic 
and slow 
magneto-acoustic waves: the latter waves have the same dispersion law as 
Alfvenic waves differing from them by polarization only.
Such degeneracy sufficiently changes the character of nonlinear interaction. 
In spite of these facts in this situation two kinds of Kolmogorov spectra are
possible with the same dependences on wave numbers as the obtained ones 
in the present paper. However, the physical motivation of existance of the two Kolmogorov 
spectra
at $\beta \gg 1$ is different.

Plan of the paper is the following. In Section 2 we introduce canonical 
description of 
an ideal MHD following to the original paper of 1972 by V.E.Zakharov and the author 
\cite{Zk} and their recent review \cite{ZK98}. By means of the 
Hamiltonian description 
in the next -- third section the average equations are derived
for  A-waves with account of interaction with slow magneto-acoustic waves. It is shown 
that from the side of 
A-waves to slow plasma motion there acts the HF force. The potential of 
this force is negative.
Therefore, unlike the interaction of Langmuir waves with ion-acoustic 
waves \cite{Z4},
plasma is drawn into regions of A-wave localization forming there density 
humps. In the same section we
analyze stability for the monochromatic A-waves.
Section 4 is devoted to the Kolmogorov spectra of weak MHD turbulence.

\section{Variational principle and canonical description}

Consider the equations of ideal MHD for barotropic flows of plasma
when the internal energy of plasma $\varepsilon$ depends only on 
its density $\rho$:
\begin{equation}
\label{1}
\frac{\partial\rho}{\partial t}+ ~\mbox{\rm div}~\rho {\bf v}=0\,;
\end{equation}
\begin{equation}
\label{2}
\frac{\partial{\bf v}}{\partial t}+
({\bf v}\nabla ) {\bf v}=
-\nabla w +\frac{1}{4\pi\rho}
~[\mbox {\rm rot}{\bf H}\times {\bf H}]\,;
\end{equation}
\begin{equation}
\label{3}
\frac{\partial {\bf H}}{\partial t}=
~\mbox {\rm rot}~[{\bf v}\times{\bf H}]\,.
\end{equation}
Here ${\bf v}$ is   plasma  velocity,  $w$  entalphy connected with
pressure $p=p(\rho)$ and internal energy $\varepsilon$ by the relations: 
$$
dw = \frac{dp}{\rho} ,\,w = \frac{\partial}{\partial\rho}\varepsilon(\rho).
$$

Let us formulate the variational principle for this system.

First of all, it should be noticed that as follows
from the equations  (\ref{1},\ref{3}) the vector
${\bf H}/\rho$ is advected by fluid particles, by another words,
each magnetic line moves together with its own particles.
This is the well known fact of frozenness of magnetic field into plasma
(see, e.g. \cite{LL2}). The given circumstance allows one to consider both
the magnetic field ${\bf H}$ and the density $\rho$ playing the roles of generalized 
coordinate.

 To formulate the variational principle we shall use the well known expression of the
Lagrangian of electromagnetic field and particles (fluid) within \cite{LL1}.
The  Lagrangian $L$ should be written in the MHD approximation. In particular, 
this means that one needs to  neglect by the contribution  from the electric field $E$ in $L$ in 
comparison with that
from the magnetic field since  $E\sim v/c~H\ll 1$ (here $c$ the light speed).
Secondly, we shall  account Eqs. (\ref{1}), (\ref{3}) and $\mbox 
{\rm div}~ {\bf H}=0$ 
as constraints. As a result, the Lagrangian can be written as follows
\begin{displaymath}
L= \frac{\rho{\bf v}^2}{2}-\varepsilon (\rho)-
\frac{H^2}{8\pi}+{\bf S}\cdot \Bigl ( \frac{\partial {\bf H}}{\partial t}-
~\mbox {\rm rot}~ [{\bf v\times H}]  \Bigr )
+\Phi \Bigl ( \frac{\partial \rho}{\partial t}+
~\mbox {\rm div}~\rho{\bf v} \Bigr ) +
\psi ~\mbox {\rm div}~ {\bf H}.
\end{displaymath}
Here ${\bf S}, \Phi$ and  $\psi$ are unknown Lagrange multipliers depending on ${\bf r}$ and $t$.

Next, using this expression of the $L$ we write down  the functional of action:
\begin{eqnarray}
\nonumber
I=\int L dt~d{\bf r},
\end{eqnarray}
which variations relative to the functions
${\bf v}, \rho$ and ${\bf H}$  yield the equations:
\begin{equation}
\label{4}
\rho{\bf v}=[{\bf H}\times \mbox{rot} {\bf S}]+
\rho\nabla\Phi\,,
\end{equation}
\begin{equation}
\label{5}
\frac{\partial\Phi}{\partial t}+
\Bigl ( {\bf v}\nabla \Bigr )\Phi -
\frac{v^2}{2}+w(\rho)=0\,,
\end{equation}
\begin{equation}
\label{6}
\frac{\partial {\bf S}}{\partial t}+
\frac{{\bf H}}{4\pi}-
[{\bf v}\times \mbox {\rm rot}~{\bf S}]+\nabla\psi =0\,.
\end{equation}
The first equation in this system gives the change of variables:
velocity ${\bf v}$ is expressed in terms of  new variables ${\bf S}$ and $\Phi$.
It is necessary to note that this change
of variables is ambiguous: 
it is possible to  add ${\bf S}$ the vector ${\bf S}_0$, and to 
 $\Phi$ the scalar $\Phi_0$ satisfying the equation
\begin{displaymath}
[{\bf H}\times \mbox {\rm rot}~{\bf S}_0]+
\rho\nabla\Phi_0=0.
\end{displaymath}
Two other equations in this system -- Eqs. (\ref{5}) and (\ref{6}) -- 
represent themselves
the Bernoulli equation for the potential  and the equation of motion for 
a new vector 
 ${\bf S}$ which contains unknown potential $\psi$. The potential $\psi$,
 in turn, is defined by fixing 
 gauge of the vector ${\bf S}$.  For example, for the Coulomb gauge  
($\mbox {\rm div}~{\bf S}=0$),
$\psi$ is determined up to arbitrary solution  $\psi_0$ of the Laplace 
equation
$\Delta\psi_0=0$:
\begin{displaymath}
\psi = \frac{1}{\Delta}~\mbox {\rm div}~
[{\bf v}\times~\mbox {\rm rot}~{\bf S}]+\psi_0.
\end{displaymath}
 In particular, if 
${\bf v}\to 0,~{\bf H}\to {\bf H}_0,~\rho\to\rho_0$ at infinity on $r$ 
then the value  $\psi_0$ is convenient to
be chosen so that $S\to 0$ at $r\to\infty$.
Then obviously  $\psi_0=-({\bf H}_0\cdot {\bf r})/(4\pi)$.

Now one needs to check that the system of equations  (\ref{4}-\ref{6}) 
does not contradict
to the original system of MHD equations. By plugging
(\ref{4}) into the equation of motion (\ref{2}),
after simple transformations, it is possible to verify 
that Eq. (\ref{2}) can be written in the form:
\begin{displaymath}
\nabla \Bigl ( \frac{\partial\Phi}{\partial t}+
({\bf v}\nabla )\Phi - \frac{{\bf v}^2}{2}+ w(\rho) \Bigr )
\end{displaymath}
\begin{displaymath}
+\Biggl [ \frac{{\bf H}}{\rho}\times~\mbox {\rm rot}~ \Biggl \{
\frac{\partial {\bf S}}{\partial t}+
\frac{{\bf H}}{4\pi}-
[ {\bf v}\times \mbox {\rm rot}~{\bf S} ]
\Biggr \} \Biggr ] =0.
\end{displaymath}
According to (\ref{5},\ref{6}) the obtained equation satisfies identically.
Thus, one can say that the new system  (\ref{1},\ref{3},\ref{5},\ref{6}) 
is equivalent the MHD system. 
Really, due to the formula  (\ref{4}) any solution of the system
 (\ref{1},\ref{3},\ref{5},\ref{6}) generates some solution of the 
original MHD system.
Under assumption about uniqueness of the Cauchy problem 
for the systems (\ref{1}-\ref{4})  and 
 (\ref{1},\ref{3},\ref{5},\ref{6}) opposite statement is valid also: 
 for any solution of the system  (\ref{1}-\ref{4}) is possible to put in 
the correspondence
 some class of solutions of the system  (\ref{1},\ref{3},\ref{5},\ref{6}). 
 For this case it is enough by using a set of quantities  
${\bf v},{\bf H},\rho$ at
 some moment of time  $t_0$ one to construct various sets of quantities  
${\bf S}$ and $\Phi$, 
 satisfying the formula  (\ref{4}) and to take them as initial conditions 
for the system
 (\ref{1},\ref{3},\ref{5},\ref{6}).

After that, by means of the Lagrange function, we determine the 
generalized momentum
 and construct the Hamiltonian of the system, by the standard way:
\begin{displaymath}
{\cal H}=\int \Bigl( ({\bf S}\cdot {\bf H}_t)+
\Phi\rho_t-{\it L} \Bigr) d{\bf r}=
\int \Biggl \{ \frac{\rho {\bf v}^2}{2} +
\varepsilon (\rho) + \frac {{\bf H}^2}{8\pi} -
\psi ~\mbox {\rm div}~ {\bf H} \Biggr \} d{\bf r},
\end{displaymath}
which coincides with the total energy of the system.  The equations of motion  
 (\ref{1},\ref{3},\ref{5},\ref{6}) is this case are nothing more than 
the Hamilton
 equations:
\begin{equation}
\label{7}
\frac{\partial \rho}{\partial t}=\frac{\delta {\cal H}}{\delta\Phi}, \quad
\frac{\partial\Phi}{\partial t}=-\frac{\delta {\cal H}}{\delta\rho},
\end{equation}
\begin{displaymath}
\frac{\partial{\bf H}}{\partial t}=\frac{\delta {\cal H}}{\delta{\bf S}}, \quad
\frac{\partial {\bf S}}{\partial t}=-\frac{\delta {\cal H}}{\delta {\bf H}}.
\end{displaymath}
Respectively, the variables  $(\rho,\Phi)$ and $({\bf H},{\bf S})$ are 
pairs of canonically conjugated values.
Change of variables  (\ref{4}) and canonical description (\ref{7}) 
were introduced for
ideal  MHD 
by Zakharov and the author in 1970 in the paper \cite{Zk}.

The transformation 
(\ref{4}) represents analog of the Clebsch representation in 
the ideal hydrodynamics.
Respectively, the fields ${\bf H}$ and ${\bf S}$ 
in the formula  (\ref{4}) play the same role as that of the Clebsch variables 
(about Clebsch variables see
\cite{lamb}, \cite{Dav}, and  recent review \cite{ZK98}). 
Later, in 1982, the authors of Ref \cite{FLS}came, in fact,  to the same change:
the velocity vector and magnetic field were represented through the scalar 
Clebsch variables that after simple transformations can be reduced 
to (\ref{4}).

 MHD flows describing by means of (\ref{4}) as well as flows of ideal 
fluids parameterized
 by the Clebsch variables  are related to the partial type of flows. 
 For example, for such MHD flows the topological invariant
of linkage of magnetic field and vorticity lines, the so-called cross helicity
\cite{moffatt}
$I=\int({\bf v\cdot H})d{\bf r}$ 
is identically equal to zero.

In  1995 Vladimirov and Moffatt  \cite{VM}  for ideal MHD 
found the analog of the Weber transformations:
\begin{equation}
\label{weber}
{\bf v}=u_{0k}({\bf a})\nabla a_k + \nabla\Phi +
\frac 1\rho[{\bf H}\times \mbox{rot} {\bf S}].
\end{equation}
Here ${\bf a}={\bf a(r},t)$ are  Lagrangian  markers of 
fluid particles (this mapping is inverse to 
 ${\bf r}={\bf r(a},t)$ which defines particle trajectory with the 
marker ${\bf a}$ ), ${\bf u}_{0}({\bf a})$
 is a new Lagrange invariant. 

The Weber transformation  (\ref{weber}) is the general transform 
which, in particular, contains at  ${\bf u}_{0}=0$, 
the change (\ref{4}) to what the authors of \cite{VM} did not pay attention.
It is interesting to note that the equations of motion for potentials 
$\Phi$ and ${\bf S}$ for the general transform (\ref{weber}) have the 
same form as for
 (\ref{5}) and (\ref{6}).  
If $\Phi$ and ${\bf S}$ are equal to zero at  $t=0$,
then ${\bf u}_{0}({\bf a})$  is nothing more than the initial velocity. 
 Just the first term in 
 (\ref{weber}) provides nonzero value of the topological
 invariant $I$. One should note also that this term, as for 
ideal hydrodynamics, is nonlinear 
 if one proceeds expansion over small amplitudes.
Recently, Ruban  \cite{ruban} (see also \cite{KR})  clarified the 
physical meaning 
of the vector  ${\bf S}$. 
Curl of the vector ${\bf S}$ can be expressed
through the shift  ${\bf d}$ between electron and ion (as fluid particles) 
in the point 
${\bf r}$ at the moment of time $t$ if initially their coordinates coincide:
$$
\mbox{rot}{\bf S}= \frac{e}{Mc}{\bf d}\frac{\rho({\bf r},t)}{\rho_0({\bf a})}.
$$
Here $M$, $e$ are ion mass and its charge, respectively,
$\rho_0({\bf a})$ the initial distribution of plasma density.

Introducing the canonical variables allow 
by the standard way (by means of the perturbation theory against small 
amplitudes of waves) 
both to classify and investigate all nonlinear processes. To this aim 
one needs in the expression for the velocity
 (\ref{weber}) as well as in the internal energy  to perform expansion 
in powers of the canonical variables.
  In the presence of the external homogeneous magnetic field  
${\bf H_0}$, in the linear approximation, one needs to keep the linear terms 
relative to $\Phi$ and ${\bf S}$, neglecting by the first (nonlinear) term in 
(\ref{weber}). 
As the result, in the velocity expansion   
\begin{equation}
\label{8}
{\bf v}={\bf v}_0+{\bf v}_1+...,
\end{equation}
the first order term is written as
\begin{displaymath}
{\bf v}_0=\frac{1}{\rho_0}[{\bf H}_0~ \times\mbox {\rm rot}~{\bf S} ]+
\nabla\Phi.
\end{displaymath}
The three independent pairs  $(\mbox {\rm div}~ {\bf H}=
\mbox {\rm div}~{\bf S}=0)$ of 
canonically conjugated quantities will correspond to three types of waves. 
In the linear approximation,
obviously, these waves will not interact. 
Their dispersion laws and polarizations 
can be found from analysis of the quadratic (relative to the canonical 
variables) Hamiltonian
${\cal H}_0$. Three-wave interaction will correspond to the cubic (relative 
to the canonical variables)
term ${\cal H}_3$. Its value will be defined from the quadratic (against wave amplitudes) 
additions to the velocity
\begin{displaymath}
{\bf v}_1=\frac{\rho_1}{\rho_0^2}[{\bf H}_0\times\mbox {\rm rot}~{\bf S} ]  +
\frac{1}{\rho_0} [ {\bf h}\times ~\mbox {\rm rot}~ {\bf S} ].
\end{displaymath} 
In this expression we take into account only 'wave'
degrees of freedom and neglect by the 
first term in 
(\ref{weber}).
Here ${\bf h}$ and $\rho_1$ are the magnetic field ${\bf H}$ and 
density $\rho$ variations from their equilibrium values ${\bf H_0}$ 
and  $\rho_0$, respectively.
As a result, the Hamiltonian represents a series with respect to 
powers of the wave amplitudes:
\begin{equation}
\label{9}
{\cal H}={\cal H}_0 + {\cal H}_{3} + ...,
\end{equation}
where the quadratic Hamiltonian is
\begin{displaymath}
{\cal H}_0=\int \Bigl \{ \frac{\rho_0{\bf v}^2_0}{2}+
\frac{{\bf h}^2}{8\pi} + c^2_s\frac{\rho_1^2}{2\rho_0}
\Bigr \} d{\bf r},
\end{displaymath}
and 
\begin{displaymath}
{\cal H}_{3} = \int \Bigl \{ \rho_0({\bf v}_0\cdot{\bf v}_1) +
\frac {\rho_1}{2} v^2_0 +qc^2_s
\frac{\rho_1^3}{2\rho^2_0} \Bigr \} dr
\end{displaymath}
is the cubic Hamiltonian.
In these formulas square of the sound speed $c_s^2$  and 
the dimensionless coefficient  $q$
appeared from expansion of the internal energy in 
powers of $\rho_1$:
$$
 \Delta\varepsilon (\rho)=
\frac{\rho_0c^2_s}{2} \Bigl \{ \Bigl ( \frac{\rho_1}{\rho_0} \Bigr )^2+
q \Bigl ( \frac{\rho_1}{\rho_0} \Bigr )^3+... \Bigr \}.
$$

 Let us now perform the Fourier transform for coordinates and
 then introduce new variables  $a_j(k)$ ($j=1,2,3$) by means 
of the following formulas
\begin{equation}
\label{10}
{\bf h}(k)={\bf e}_1(k)\sqrt {2\pi\omega_1}
(a_1(k)+a^*_1(-k)) +
\end{equation}
\begin{displaymath}
+{\bf e}_2(k) \sum\limits_{l=2,3}\lambda_l\sqrt{2\pi\omega_l}
(a_l(k)+a^*_l(-k));
\end{displaymath}
\begin{displaymath}
{\bf S}(k)=-i{\bf e}_1(k) \frac{1}{\sqrt{8\pi\omega_1}}
(a_1(k)-a^*_1(-k)) -
\end{displaymath}
\begin{displaymath}
-i{\bf e}_2(k)\sum\limits_{l=2,3}\lambda_l\frac{1}{\sqrt{8\pi\omega_l}}
(a_l(k)-a^*_{l}(-k));
\end{displaymath}
\begin{displaymath}
\rho_1(k) =\sum\limits_{l=2,3}
\Biggl ( \frac{\rho_0\omega_l}{2c^2s} \Biggr )^{1/2} \mu_l
(a_k(l)+a^*_{-k}(l))
\end{displaymath}
\begin{displaymath}
\Phi(k)=-i\sum\limits_{l=2,3}
\Biggl ( \frac{c^2_s}{2\rho_0\omega_l} \Biggr )^{1/2} \mu_l
(a_l(k)-a^*_l(-k)).
\end{displaymath}
Here
\begin{displaymath}
\omega_1(k) = |({\bf k\cdot V}_A)|\,,
\end{displaymath}
\begin{displaymath}
\omega_{2,3}(k)=\frac{1}{2} \left|\sqrt {k^2V^2_A+k^2c^2_s+
2({\bf k\cdot V}_A)kc_s}\pm
 \sqrt {k^2V^2_A+k^2c^2_s-2({\bf k\cdot V}_A)kc_s} \right|
\end{displaymath}
are the dispersion laws of Alfvenic waves (1), fast (2) and slow (3)
magneto-acoustic waves, respectively;
$$
{\bf e}_1(k)=
\frac{[{\bf k\times n}_0]({\bf k\cdot n}_0)}{|[{\bf k\times n_0]}||
({\bf k \cdot n}_0)|}, \qquad
{\bf e}_2(k)=
\frac{[{\bf k}\times [{\bf k\times n}_0]]}
{k|[{\bf k\times n}_0]|}\,\, 
$$
are corresponding to them the unit polarization vectors; 
${\bf n}_0={\bf H}_0/H_0$ the unit vector
along mean magnetic field;
${\bf V}_{A}={\bf H}_0/(4\pi\rho_0)^{1/2}$ the Alfven velocity;
\begin{displaymath}
\lambda_2=-\mu_3=- \Biggl ( 1 -
\frac{\omega^2_3-k^2c^2_s}{\omega^2_2-k^2c^2_s}
\Biggr )^{1/2}\,,
\end{displaymath}
\begin{displaymath}
\lambda_3=\mu_2= \Biggl ( 1 -
\frac{\omega^2_2-k^2c^2_s}{\omega^2_3-k^2c^2_s}
\Biggr )^{-1/2}\,.
\end{displaymath}
The given change to the new  variables  $a_k(j)$ represents the canonical 
$U-V$ transformation diagonalizing the Hamiltonian  ${\cal H}_0$:
\begin{displaymath}
{\cal H}_0=\sum_j\int \omega_j(k)a_j(k)a_j^*(k)d{\bf k}\,.
\end{displaymath}
In this case the amplitudes  $a_j(k)$ have the meaning  of normal variables,
and, respectively, the equation of motion have the standard canonical form:
\begin{displaymath}
\frac{\partial a_j(k)}{\partial t}=-
i\frac{\delta {\cal H}}{\delta a^*_j(k)}
\end{displaymath}
In the linear approximation  $a_j(k)$ obey the equations:
$$
\frac{\partial a_j(k)}{\partial t} + i\omega_j(k)a_j(k)=0\,,
$$
namely, with time the amplitude modulus   $|a_i(k)|$ does not change, but
the phase increases linearly with $t$.

In order to find the expression for the Hamiltonian of interaction 
in terms of the $a_j(k)$ variables one needs to substitute transformation 
(\ref{10}) into (\ref{9}). As a result, the Hamiltonian of interaction 
will represent the integro-power series relative to theses variables. 
In the lowest order with respect to the wave amplitudes the main nonlinear
process will be the resonant three-wave interaction. The corresponding 
Hamiltonian is equal to 
\begin{equation} \label{int}
{\cal H}_{int}=\frac 12\int\sum_{lmn} \Bigl [ V^{lmn}_{kk_1k_2}
a^*_l(k)a_m(k_1)a_n(k_2)+ c.c. \Bigr ]
\delta_{k-k_1-k_2}dkdk_1dk_2.
\end{equation}
This Hamiltonian can be obtained after substitution of the transform 
(\ref{10}) into the cubic Hamiltonian ${\cal H}_{3}$ and subsequent extraction
from there the resonant terms. The rest terms in ${\cal H}_{3}$ are small: 
they can be excluded by means of the canonical transformation (for details 
see the review \cite{ZK98}).
One should note that calculation of matrix elements $V^{ije}_{kk_1k_2}$ 
in this scheme
is a pure algebraic procedure  requiring performance of the Fourier 
transform in all integrals,
substitution (\ref{10}) and forthcoming symmetrization with respect to 
the $a_k(i)$ variables,
for instance, in  (\ref{int}) against pairs  $(k_1,m)$ and $(k_2,n)$.

\section{Average equations}

Expressions for dispersion laws and matrix elements
of interaction can be sufficiently simplified 
for plasma with low  $\beta=8\pi nT/H^2 $ (it is ratio of thermal plasma 
pressure   $nT$ 
and magnetic field pressure $ H^2/8\pi$. The condition $\beta\ll 1$ means 
that  $V_A\gg c_s$.
In this limit fast magneto-acoustic waves have isotropic dispersion law 
$\omega_2=kV_A$ and their phase (as well as group) velocity coincides with
the value of the group velocity of Alfvenic waves. In the linear approximation 
the plasma velocity in Alfvenic and fast magneto-acoustic waves 
is given by the expression
\begin{displaymath}
{\bf v}_{HF}=\frac{1}{\rho_0}[{\bf H}_0\times\mbox {\rm rot}~{\bf S} ].
\end{displaymath} 
The potential part of the velocity $\nabla\Phi$ occurs to be small due to 
smallness of the parameter $\beta$.
For slow magneto-acoustic waves, on the contrary,  the main contribution 
is given by the potential part -- the velocity turns out to be directed 
along the magnetic field  
${\bf H_0}$:
\begin{equation}
\label{sound}
{\bf v}_{s}={\bf n_0}\frac{\partial \Phi}{\partial z},
\end{equation}
and the dispersion law for slow magneto-acoustic waves becomes 
strongly anisotropic:
\begin{equation}
\label{sound1}
\omega_3\equiv \Omega_s=|k_z|c_s.
\end{equation}
The transverse components of the velocity for these waves
 $[{\bf H}_0\times\mbox {\rm rot}~{\bf S} ]/{\rho_0}$ are compensated by 
 $\nabla_{\perp}\Phi$.

If plasma is collisionless and strongly isothermal ($T_e\gg T_i$), 
the slow magneto-acoustic waves
represent themselves magnetized ion-acoustic waves
(for details see  \cite{kuz}). In this case in (\ref{sound1}) 
$c_s=\sqrt{T_e/M}$. 

As far as the nonlinear interaction of the MHD waves concerns,
for strongly magnetized plasma the main nonlinear process is the process of 
scattering  of the Alfvenic and fast magneto-acoustic waves on 
the slow magneto-acoustic waves (that can be verified directly by 
comparing the computing matrix elements $V^{lmn}$ for (\ref{int})).
In this process the former waves (further we shall call them as A-waves)
plays the role of the high-frequency (HF) waves relative to the latter ones
(these waves shall be simply called  sound or $S$-waves).
This conjecture follows directly from the analysis of the resonant conditions 
for the given type of decay:
\begin{equation}
\label{decay}
\omega_A(k)=\omega_A(k_1)+\Omega_s(k_2),\,\,
{\bf k}={\bf k_1}+{\bf k_2}.
\end{equation}
Qualitatively it is easily to understand how this interaction looks like. 
While propagation of the packet of A-waves the mean characteristics of plasma
(its density and mean velocity) due to the action of A-waves will be slowly varied.
By this reason the mean Alfven velocity will differ from its local value by
the quantity $\Delta V_A =- V_A\rho_{1s}/(2\rho_0)$ where
$\rho_{1s}$ is low-frequency (LF) density variation. It results in the 
frequency addition
of A-waves  $\Delta\omega_{\rho}\sim k \Delta V_A$.
  Due to slow motion of plasma with the drift velocity 
 $v_D$ the frequency of A-waves changes at
$\Delta\omega_D\sim kv_D$ . 
The ratio of these two additions, 
$\Delta\omega_D$ and $\Delta\omega_{\rho}$,however, occurs to be small: 
 $\sim c_s/V_A$.  Thus, the main interaction is the 
scattering on the LF density fluctuations. At the same time the LF plasma 
characteristics
will be changed due to the action of the HF force induced by A-waves.

 The most simple way to find the expression of the HF force is to perform
 average of the Hamiltonian over the HF oscillations. The result of this 
average is
 the following 
\begin{equation}
\label{H}
{\cal H}={\cal H}_0 +{\cal H}_{\mbox {\rm int}},
\end{equation}
where
$$
{\cal H}_0 = \int\left \{ \frac{1}{2\rho_0}\langle [ {\bf H}_0 
\times\mbox {\rm rot}~ {\bf S} ]^2 \rangle
+ \frac{\langle {\bf h}^2\rangle}{8\pi} \right \} d{\bf r}\\
+\int\left \{ \frac{\rho_0\Phi_z^2}{2}+c^2_s\frac{\rho_{1s}^2}
{2\rho_0}  \right\} d{\bf r}\,,
$$
$$
{\cal H}_{\mbox {\rm int}}=
=-\int \frac{\rho_{1s}}{2\rho^2_0}
\langle [ {\bf H}_0 \times \mbox {\rm rot}~ {\bf S} ]^2 \rangle d{\bf r}.
$$
Here angle brackets mean average over high frequency. 
The first integral in ${\cal H}_0$
corresponds to (linear) A-waves, the second one describes 
magnetized acoustic waves and the last term
is responsible for nonlinear interaction.
Variation of the Hamiltonian of interaction on  
 $\rho_{1s}$ gives the expression for the HF potential :
\begin{equation}
\label{U}
U\equiv M\frac{\delta {\cal H}_{int}}{\delta\rho_{1s}}=- \frac{1}{2Mn^2_0}
\langle [ {\bf H}_0 \times\mbox {\rm rot}~ {\bf S} ]^2 \rangle.
\end{equation}
In according with (\ref{U}) the equation of motion 
for the potential  $\Phi_s$ takes the form:
\begin{equation}
\label{phi}
\frac{\partial\Phi_s}{\partial t} +
c^2_S\frac{\rho_{1s}}{\rho_0}=
\frac{\langle [ {\bf H}_0 \times\mbox {\rm rot}~ {\bf S} ]^2 \rangle}
{2\rho^2_0}.
\end{equation}
It is important to notice that the HF potential (\ref{U}) is negative.
This means that in the region of localization of A-waves
the HF force will form, instead of density wells,
as it is for the interaction between Langmuir and ion-acoustic waves (see \cite{Z3}),
the density humps.

The equations of motion are closed by the continuity equation
for $\rho_{1s}$ which, in accordance with  (\ref{sound}),  has the form:
\begin{equation}
\label{rho}
\frac {\partial\rho_{1s}}{\partial t} + \rho_{0}\frac{\partial^2 \Phi_s}
{\partial z^2}=0.
\end{equation}

From (\ref{phi}) and (\ref{rho}) we have 
\begin{equation}
\label{rho1}
\frac {\partial^2\rho_{1s}}{\partial t^2} - c_s^2\frac{\partial^2\rho_{1s}}
{\partial z^2}=-\frac {1}{2\rho_0}
\frac{\partial^2}{\partial z^2}{\langle [ {\bf H}_0\times ~\mbox {\rm rot}~ 
{\bf S} ]^2 \rangle}.
\end{equation}

To write the equation for  A-waves one needs to make average explicitly 
in the Hamiltonian of interaction ${\cal H}_{int}$.
It corresponds to keeping in  ${\cal H}_{int}$ terms containing products  
$a^*_{\lambda}
a_{\lambda_1}$ where the index $\lambda=1,2$ enumerate the HF waves:
\begin{displaymath}
H_{\mbox {\rm int}} =-\int \frac{\rho_{1s}(\kappa )}{2\rho_0}
\sum\limits_{\lambda\lambda_1}
F^{\lambda\lambda_1}_{kk_1}
a^*_{\lambda}(k)
a_{\lambda_1}(k_1)
\delta_{k-k_1-\kappa}
dkdkd\kappa.
\end{displaymath}
Here
\begin{displaymath}
F^{\lambda\lambda_1}_{kk_1}=
(\omega_{\lambda} (k)\omega_{\lambda_1}(k_1))^{1/2}
({\bf n}_{\lambda}(k)\cdot{\bf n}_{\lambda_1}(k_1)), \quad
{\bf n}_2=\frac{{\bf k}_{\bot}}{k_{\bot}}, \quad
{\bf n}_1=-[ {\bf n}_2 {\bf n}_0 ]
\end{displaymath}
As the result the equations for A-waves have the form:
\begin{equation}
\label{14}
\frac{\partial a_{\lambda}(k)}{\partial t} + i\omega_{\lambda}(k)
a_{\lambda} (k)=
-i\frac{\delta H_{int}}{\delta a^*_{\lambda}(k)}; \quad
\lambda = 1,2.
\end{equation}

For isothermal collisionless plasma  ($T_e\approx T_i$) slow 
magneto-acoustic waves are
absent due to strong Landau damping on ions. 
Correspondingly, the decay interaction of A-waves transforms into the 
induced scattering on ions. 
In this case  the equations 
(\ref{rho1}) have to be changed by the drift kinetic equation \cite{vvS} for slow 
variation of distribution
functions of ions  $f_i$ (compare with \cite{K76}):
\begin{equation}
\label{12}
\frac{\partial f_i}{\partial t}+
v_z\frac{\partial f_i}{\partial z}-
\frac{1}{M} \frac{\partial}{\partial z}(e\tilde\varphi +U)
\cdot\frac{\partial f_0}{\partial v_z}=0,
\end{equation}
together with the quasi-neutrality condition for slow motions 
$(\Omega_k=k_zc_S\ll \omega_{pi})$
\begin{equation}
\label{13}
\delta n_i = \int f_id{\bf v} = \frac{n_0}{T_e}
e\tilde\varphi  = \frac{\rho_{1s}}{M},
\end{equation}
where $f_0$ is the equilibrium distribution function of ions,
and 
$\tilde\varphi$ the LF electrostatic potential.
In this case the equation of motion for A-waves retrains 
the form of  (\ref{14}), and 
the density is expressed linearly through the HF potential 
by means of the Green function for the system (\ref{12},\ref{13}):
\begin{equation}
\label{G}
G_{\kappa\Omega} \equiv \frac{\rho_{1s}(\kappa,\Omega)}{U_{\kappa\Omega}} 
=-\frac{n_0\kappa^2}{\omega_{pi}^2}\frac{\epsilon_e
\epsilon_i}{\epsilon_e+\epsilon_i}.
\end{equation}
Here  $\rho_{1s}(\kappa,\Omega)$ and $U_{\kappa\Omega}$ are 
the Fourier images of the LF density and 
the HF potential, respectively, 
 $\epsilon_{e,i}$ partial dielectric constants of electrons 
and ions which are equal:
$$
\epsilon_{e}= \frac{1}{\kappa^2 r_d^2} \,,
$$
$$
\epsilon_{i}= \frac{4\pi e^2}{M\kappa^2}\int \frac{\kappa_z({\partial f_0}/
{\partial v_z})}{\Omega-\kappa_z v_z}d{\bf v}\,,
$$
where $r_d^2={T_e}/({4\pi n_0e^2})$ is square of the Debye radius.

In non-isothermal plasma  ($T_e\gg T_i$) the Green function (\ref{G}) 
transforms into
$$
G_{\kappa\Omega} = \frac{n_0\kappa^2_z}{\Omega^2-\kappa^2_zc^2_s},
$$
that coincides with the expression given by the equation (\ref{rho1}). 

The system of equations  (\ref{12})-(\ref{G}) completely describes 
interaction of A-waves in magnetized plasma with arbitrary ratio 
of ion and electron temperatures. In this case, however, the Hamiltonian
$H_0+H_{\mbox {\rm int}}$ is not conservative quality due to the 
Landau damping on ions.

\section{Instability of monochromatic wave}

Let us now analyze the obtained equations. 
We start from study of dynamics of the narrow packet of A-waves.

A qualitative understanding about this process can be obtained
from stability analysis of monochromatic A-wave. In the sake of simplicity
we shall restrict by consideration of stability of Alfvenic wave 
in the hydrodynamic limit.
For collisionless plasma the latter assumes that the phase velocity  
of bending
$\Omega /\kappa_z$ for A-waves exceeds the thermal ion velocity $v_{Ti}$. In this case
for slow motion one can neglect by the Landau damping on ions and use
the equations  (\ref{rho1}) or (\ref{G}).
 One should remember that in strongly non-isothermal plasma plasma
 the sound waves are the  eigen oscillations but at 
the same time in plasma with 
 $T_e\approx T_i$ sound waves represent the induced oscillations of 
the plasma density.
However, at  $\Omega /\kappa_z\gg v_{Ti}$ the hydrodynamic description 
can be applied in both cases.
 
Then it is convenient to express  $\rho_{1s}$ through the 
normal variables  $a_3(k)\equiv b_k$:
\begin{displaymath}
\rho_{1s}(k)=\Bigl ( \frac{\rho_0\Omega_k}{2c_s} \Bigr )^{1/2}
(b_k+b^* _{-k}).
\end{displaymath}
The equations of motion for $b(k)$ are obtained from variation of 
of the full Hamiltonian  $H_0+H_{\mbox {\rm int}}$:
\begin{equation}
\label{15}
\frac{\partial b_k}{\partial t}+i\Omega(k)b_k =
-i~\delta H_{\mbox {\rm int}}/\delta b^*_k.
\end{equation}
In the equations  (\ref{14}), (\ref{15})  the solution
\begin{displaymath}
a_{\lambda}(k) =\frac{A}{\omega^{1/2}_0}
\delta_{\lambda 1}e^{-i\omega_0t}
\delta_{k-k_0}, \quad\quad
b_k=0,  \quad\quad
\omega_0=\omega_1(k_0)
\end{displaymath}
corresponds to the monochromatic Alfvenic wave.
The amplitude of Alfvenic wave here is chosen by such a way so that
$|A|^2$ coincides with the energy density of oscillations $W$. 

Linearizing of the equations  (\ref{12})-(\ref{G}) on the background of exact 
solution and assuming for perturbations:
\begin{displaymath}
\delta a_{\lambda}(k)\sim e^{-i(\Omega +\omega_0)t}
\delta_{k-k_0-\kappa},
\end{displaymath}
\begin{displaymath}
\delta a_{\lambda}^*(k)\sim e^{-i(\Omega -\omega_0)t}
\delta_{k-k_0+\kappa},
\end{displaymath}
for $\Omega$ we have the following dispersion relation:
\begin{equation}
\label{16}
\frac{WG}{4Mn^2_0\omega_0} \sum\limits_{\lambda}
\Biggl \{  \frac{|F^{1\lambda}_{k_0,k_0+\kappa} |^2}
{\Omega + \omega_0-\omega_{\lambda} (k_0+\kappa)} +
\frac{|F^{1\lambda}_{k_0,k_0-\kappa} |^2}
{-\Omega + \omega_0-\omega_{\lambda}(k_0-\kappa)}
\Biggr \} =1\,.
\end{equation}

We shall present now results of investigations of the dispersion equation
(\ref{16}) in the different cases in dependence on the energy wave density 
 $W$ and on the temperature ratio. 

At $T_e\gg T_i$ and sufficiently small amplitudes
the decay instability takes place with generation of ion magnetized sound 
\cite{GO}. For this instability the eigen  frequency  $\Omega$ 
is expressed through the matrix element of the decay interaction 
\begin{equation}
\label{matrix}
V^{\lambda\lambda_1}_{kk_1k_2}=
\Biggl  ( \frac{\Omega(k_2)}{8\rho_0c^2_s} \Biggr )^{1/2}
F^{\lambda\lambda_1}_{kk_1}
\end{equation}
and the quantity  $W$:
\begin{equation}
\label{omega}
\Omega=\frac 12 \left[\omega_0 -\omega_{\lambda}(k_0-\kappa) +
\Omega(\kappa)\right] \pm
\left\{\frac 14\left[\omega_0 -\omega_{\lambda}(k_0-\kappa) -
\Omega(\kappa)\right]^2
-\frac{W}{\omega_0}|V^{1\lambda}_{k_0,k_0-\kappa,\kappa}|^2\right\}^{1/2}.
\end{equation} 
Hence it follows that the instability 
takes place near the resonant surface
\begin{equation}
\label{sur}
\omega_{0} = \omega_{\lambda}(k_0-\kappa)+ \Omega(\kappa)
\end{equation}
with maximum of the growth rate 
\begin{equation}
\label{growth}
\Gamma = \Bigl [ \frac{W}{8nT} \frac{\Omega_{\kappa}}{\omega_0}
| F^{1~\lambda}_{k_0~k_0-\kappa} |^2 \Bigr ]^{1/2}.
\end{equation}
The growth rate width as a function of frequency
occurs to be of the order of magnitude of the maximal
growth rate (\ref{growth}).  

Because the matrix element is proportional to 
the frequency of slow sound
the maximum value of the growth rate on the resonant surface 
 (\ref{sur}) is attained at the maximal value of 
$|\kappa_z|$. For decay on Alfvenic wave and slow sound  
$\max|\kappa_z|\approx 2|k_{0z}|$, so that the secondary Alfvenic wave
propagates in the opposite direction to the pumping Alfvenic wave.
Such behavior of the decay instability is typical for the Mandelstamm-Briullien
scattering, the matrix element of which is proportional to square root 
from the sound 
momentum transfered by scattering light. For light such dependence  provides 
the maximal back scattering. 

It is not difficult to investigate the decay instability for all 
other possible channels of decay
$A\to A+S$. The growth rate in all these cases are of the same order 
of magnitude as (\ref{growth}):
 $$
 \Gamma\sim (\omega_0\Omega_sW/nT)^{1/2}.
 $$
This instability takes place at 
$$
W/nT < \beta^{1/2}.
$$
With increase of  $W$ the decay instability is transformed.
At $W/nT > \beta^{1/2}$ in the dispersion relation 
(\ref{16}) it is possible to neglect by 
$\Omega^2_s$ against $\Omega^2$. 
Then the unstable wave vectors will lie on the surface  
$\omega_1(k_0)=\omega_{\lambda}(k_0-\kappa)$. This instability is called 
as the modified decay instability \cite{Z4,ZR}. For the interaction of 
Alfvenic waves and slow sound this instability has the growth rate maximal at 
$\kappa_z=2k_{0z}$:
\begin{equation}
\label{17}
\Gamma \approx \frac{\sqrt{3}}{2}\omega_0 \Bigl ( \frac{W}
{\rho_0V_A^2} \Bigr )^{1/3}.
\end{equation}
Value of this growth rate does not depend on temperature and 
therefore this instability takes place at $W/nT > 1$ up to the values
$\beta^{-1}$ when the main approximation - adiabaticity approximation - 
looses its applicability: $\Gamma \sim \omega_0$.

 For another channels the instability with growth of
$W/nT$ has the same character: at $W/nT > \beta^{1/2}$
the growth rate is maximal in the region  $\kappa\sim k_0$  and 
has the same order of magnitudes  as  (\ref{17}).  

The decay instability  (\ref{growth}) for arbitrary channel  
$A\rightarrow A+S$, as it is easily seen, relates to the 
convective type of instabilities. The excited waves,
according to (\ref{omega}),  have the group velocities 
 strongly different from the group velocity of the pumping wave.
Therefore for the wave packet with the characteristic scale $L$ 
this instability will be essential only for large enough lengths $L$
when the amplification coefficient $G$ exceeds a value of the Coulomb logarithm 
$\Lambda$:
$$
G=\Gamma L/V_A \approx \Lambda.
$$ 
For less lengths $L$, the decay instability is not important:
during propagation of perturbation of through the whole packet  
perturbations amplify for a small value. In this case 
dynamics of the packet will be defined by slow
processes. Among them the most important ones are 
such processes for which unstable perturbations propagate 
together with the wave packet. If it is a decay instability, 
then it has to be absolute 
(in the frame moving together with the packet). In particular, 
this is one of the reasons 
of appearance of collapse for fast magneto-acoustic waves and 
of affect of sound collapse on the fine structure of collisionless
shocks in plasma \cite{KMS,KM}.
Collapse  of fast magneto-acoustic waves appears due to three-wave interaction 
in which only fast magneto-acoustic waves take part.

\section{Kolmogorov spectra}

In the previous section we considered the stability problem for narrow 
in $k$-space
wave packet. In this case for decay of monochromatic wave under the resonant conditions 
 (\ref{sur}) (namely, for the maximal value of the growth rate (\ref{growth}))
sum of the phases of exciting waves  $\phi_A$ and $\phi_s$ are strongly 
connected with
the phase of the pumping wave $\phi_0$:
$$
\phi_0+\pi/2=\phi_A+\phi_s.
$$
(It is easily to check that this phase correlation 
is lost with leaving the resonance (\ref{sur}).) Simultaneously,  
a difference in  phases for the pair of exciting waves with fixed 
 $\kappa$ remains arbitrary. Both these factors introduce to 
 the system of interacted triads, connected with the pumping 
 wave an element of randomness.  Thus, each triad is characterized 
by one random phase. 
 At the next step - at the second cascade a new random phases are 
 added so that a memory about the pumping wave will be lost.  For 
multiple repetition of of this process
 the system of waves must transform in the turbulent state when the phases 
of waves can be considered
 random. Therefore the stochastization time should be a few inverse 
growth rate (\ref{growth}).

Such scenarion of transition to turbulence seems to be sufficiently plausible. 
Now there are being performed a number of numerical experiments to check 
this hypothesis
(see, for instance, \cite{ZP},\cite{ZDV}).

Hence it becomes clear that the regime of developed wave turbulence 
should be characterized by the wide spectrum of waves. For small enough 
intensities of waves it is enough one to restrict by consideration 
of the pair correlation functions only which obey the kinetic equations 
for waves. 
This regime is called weak turbulent. 

In the case of the weak MHD turbulence  at $\beta \ll 1$  we have three pair 
correlation 
functions defined by the following formulas:
\begin{displaymath}
\langle a_{\lambda}(k) a^*_{\lambda_1} (k_1)\rangle=
N_{k}^{\lambda}\delta_{\lambda\lambda_1}\delta_{k-k_1},\,\,
 \langle b_kb^*_{k_1} \rangle =
n_k\delta_{k-k_1}
\end{displaymath}
where the quantities $N_{k}^{\lambda}$, $n_k$,  having a meaning of the 
occupation numbers, satisfy the
following system of kinetic equations:
\begin{equation}
\label{18}
\dot n_k=2\pi\int |V_{k_1k_2k} |^2
(N_{k_1}N_{k_2}-n_kN_{k_1}+n_kN_{k_2})
\delta_{k+k_1-k_2}\delta_{\Omega+\omega_{1}-
\omega_{2}}dk_1dk_2,
\end{equation}
\begin{equation}\label{181}
\dot N_k=2\pi \int |V_{kk_1k_2} |^2
(N_{k_1}n_{k_2}-N_kn_{k_2}-N_kN_{k_1})
\delta_{k-k_1-k_2}
\delta_{\omega-\omega_{1}-\Omega_{2}} dk_1dk_2 
\end{equation}
\begin{displaymath}
- 2\pi \int |V_{k_1kk_2} |^2 (N_kn_{k_2}-N_{k_1}n_{k_2}-N_kN_{k_1})
\delta_{k_1-k-k_2}\delta_{\omega_{1}-\omega-\Omega_{2}} dk_1dk_2.
\end{displaymath}
Here $\omega \equiv \omega(k)$, $\omega_1\equiv \omega(k_1)$, and  so on. 
In these equations (as well as
below) we omit summation over $\lambda$. In order to include 
it  one needs to change  $dk_1\to \sum\limits_n dk_1, ~
N_k\to N^{\lambda}_k,~ \omega_k\to  \omega_{k\lambda}$, $V_{kk_1k_2}\to 
V^{\lambda\lambda_1}_{kk_1k_2}$  and so on.

The equations  (\ref{18},\ref{181}) assume weakness 
of the nonlinear interaction between waves.
In this concrete case the most essential criterion is 
 $$
\Omega_s\gg 1/\tau,
$$
where $\tau $ is the characteristic nonlinear time 
defined by the kinetic equations 
(\ref{18},\ref{181}).   To estimate the value 
of $\tau $ one needs to take into account that in each act of 
decay and inverse process - merging of waves,  
the frequencies of A-waves change for the small value 
$\Delta\omega_A = \Omega_s\ll \omega_A$, 
namely, the energy transfer of A-waves along spectrum  
has a diffusive character. Due to this fact, we have the following 
estimate for 
$\tau $:
$$
\frac{1}{\tau}\sim \omega_A\frac{W}{\rho V_A^2}.
$$
Notice, that this value for $\tau $ exceeds significantly the 
stochastization time, defined by the inverse
growth rate  (\ref{growth}) $\Gamma^{-1}$.  

Hence we have finally the criterion:
$$
\frac{W}{\rho V_A^2}\ll \beta^{1/2}.
$$

Next, let us include into the kinetic equations 
(\ref{18}), (\ref{181}) both sources of turbulence and its 
dissipation. For this aim in the left hand sides of the equations
we introduce new  terms $\Gamma_kn_k$ and
$\gamma_{k\lambda}N_{k\lambda}$, respectively. We suppose here that 
the pumping region 
 ($\Gamma_k,
\gamma_{k\lambda}>0$) and dissipation region ($\Gamma_k,
\gamma_{k\lambda}<0$) are well separated in $k$-space by the intermediate 
region - the inertial interval, where dynamics of turbulence is 
defined by nonlinear interaction 
between waves only. In the inertial interval we shall neglect 
by influence of both pumping and dissipation (that is necessary to be proved)
then  distributions $n_k$ and $N_{k\lambda}$ do not depend on the 
concrete form of $\gamma_k$  and $\Gamma_k$.

We would like to remind  that in the theory of developed hydrodynamic 
turbulence to determine turbulent spectrum - distribution of 
energy for velocity fluctuations it is enough to use two hypothesizes of 
A.N.Kolmogorov 
\cite{kol}.
 The first hypothesis 
about self-similarity says that the spectrum of turbulence is defined 
by the unique
quantity $P$ -- a constant flux of energy along scales (from large to small ones 
where dissipation due to viscosity becomes essential).
The second hypothesis assumes that the interaction of fluctuations with 
different scales has a
local character.

If one applies these hypothesizes to the given case 
then spectra of turbulence in the inertial interval 
can be found by means of the dimensional analysis. 
In the given situation the kinetic equations 
(\ref{18}), (\ref{181}) have two conservation laws: conservation of the total energy and 
the total number of HF waves. To each integral of motion there should correspond
its proper Kolmogorov spectrum. 
So, for a constant flux of the number of HF waves  
$N_k^{\lambda}$ 
\begin{displaymath}
P_N=\frac{\partial}{\partial t} \sum\limits_{\lambda}
\int N_{k\lambda} dk\,,
\end{displaymath}
we have the spectrum:
\begin{equation}
\label{19}
N_{k\lambda}\sim P^{1/2}_Nk^{-4},  \qquad
n_k\sim P^{1/2}_Nk^{-4}.
\end{equation}
For a constant flux of energy 
\begin{displaymath}
P_{\varepsilon}=\frac{\partial}{\partial t}
\int (\omega_kn_k+
\sum\limits_{\lambda}\omega_{k\lambda}N_{k\lambda})dk\,,
\end{displaymath}
one can get the estimate:
\begin{equation}
\label{19'}
N_{k\lambda}\sim P^{1/2}_{\varepsilon}k^{-3/2}, \qquad
n_k\sim P^{1/2}_{\varepsilon}k^{-3/2}\,.
\end{equation}
From conservation in the inertial range of the total number of HF waves and the 
energy it is easily to establish that the flux of HF particles is 
directed towards small $k$ region, 
and energy flux is directed to the short wave region.

These -- sufficiently rough --  estimations for the spectra (\ref{19}), 
(\ref{19'}) 
can pretend only to the right dependence on both wave numbers and fluxes, 
but they  don't account the diffusive character of decays. 
Notice also that these estimates are significantly based on an assumption of 
the interaction locality. 

The spectra (\ref{19}) and (\ref{19'})  don't account also fine properties 
of distribution functions --
their angle dependences, namely, they are defined up to 
arbitrary functions of angles. 
To find these dependences one needs to solve the exact 
equations (\ref{18}), (\ref{181}). 
It turns out that solution of these equations can be found for interaction 
of Alfvenic and slow 
acoustic waves $(N_2\equiv 0)$.  For this case it is convenient
to represent the equations  (\ref{18}), (\ref{181}) in the form:
\begin{equation}
\label{21}
\dot n_k=-\int U_{k_2 | kk_1} T_{k_2 | kk_1}dk_1dk_2\,, 
\end{equation}
\begin{equation}
\label{21'}
\dot N_k=\int ( U_{k|k_1k_2} T_{k| k_1k_2} -
U_{k_1|kk_2} T_{k_1|kk_2})
dk_1dk_2\,,
\end{equation}
where the following notations are introduced:
\begin{displaymath}
U_{k|k_1k_2}=
2\pi | V^{11}_{k'k'_1k_2} |^2
\delta_{k-k_1-k_2}\delta_{\omega -\omega_1-\omega_2}\,,
\end{displaymath}
\begin{displaymath}
T_{k|k_1k_2}=N_{k_1}n_{k_2}-
N_kn_{k_2} -
N_kN_{k_1}.
\end{displaymath}
It is easily to see that the equations  (\ref{21}), (\ref{21'}) 
have the thermodynamic equilibrium 
solutions:
\begin{displaymath}
N_k=\frac{N}{\omega_k+\mu}, \qquad
n_k=\frac{T}{\Omega_k}
\end{displaymath}
-- the Rayleigh-Jeans distributions which annulate the collision terms.

To find the non-equilibrium distributions it should be noted that the function
$U$ has the following properties. (i) $U$ is bi-homogeneous function 
of its arguments $k_{z}$ and $k_{\bot}$ with homogeneity 
degrees, respectively, equal to 
$+1$ for $k_z$ and $-2$ for $k_{\bot}$. This means that if one performs stretching 
of $k_z$, $k_{1z}$, $k_{2z}$
in $\lambda$ times then $U$ multiplies in $\lambda^{+1}$ times: 
$U\to\lambda^{+1} U $. If 
one makes the similar transform for all $k_{\bot}$: 
$k_{\bot}\to \mu k_{\bot}$,  then $U$ 
transforms as $U\to \mu^{-2}U$. Besides, (ii) $U$ is invariant 
with respect to rotation around
$z$-axis - the direction of mean magnetic field ${\bf H}_0$. 

 Due to these properties  solution is naturally sought in the form:
\begin{equation}
\label{22}
n_k=Ak^{\alpha}_zk^{\beta}_{\bot}, \qquad
N_k=Bk^{\alpha}_zk^{\beta}_{\bot}.
\end{equation}

Consider the stationary equation (\ref{21'}):
\begin{equation}
\label{23}
\int ( U_{k|k_1k_2} T_{k| k_1k_2} -
U_{k_1|kk_2} T_{k_1|kk_2}) =0.
\end{equation}
Let us make a mapping of the integration area of the second integral (which is 
given  by its resonance conditions -- $\delta$-functions) to the integration 
area of the first integral. For this aim it is convenient to 
introduce the complex variables
$ \zeta= k_x + ik_y$. Then the integration area defined by the 
corresponding 
conservation laws
\begin{displaymath}
k_{z1} - k_z - k_{z2} =0,
\end{displaymath}
\begin{displaymath}
\zeta_1 - \zeta - \zeta_2 = 0,
\end{displaymath}
\begin{displaymath}
\omega_1 - \omega - \Omega_2 = 0,
\end{displaymath}
with the help of the mapping relative to all $k_z$ and $\zeta$:
\begin{equation}
\label{24}
k_z = k_z' \frac{k_z}{k_z'},  \qquad \zeta =\zeta' \frac{\zeta}{\zeta'},
\end{equation}
\begin{displaymath}
k_{z1} = k_z \frac{k_z}{k_z'},  \qquad \zeta _1=\zeta \frac{\zeta}{\zeta'},
\end{displaymath}
\begin{displaymath}
k_{z2} = k_z'' \frac{k_z}{k_z'},  \qquad \zeta_2=\zeta'' \frac{\zeta}{\zeta'},
\end{displaymath}
transforms into the integration area of the first integral in (\ref{23}).
Each such transformation (separately with respect to $k_z$ and $\zeta$)
represents itself the operation of inversion: for 
$z$-components of wave vectors against the point $k_z$,  
and for transverse components relative to the circle with radius
 $|k_{\bot}|$.  Under this mapping the vector 
${\bf k}$ transforms into ${\bf k_1}$, 
${\bf k_1}$  into $ {\bf k}$ and  ${\bf k_2}$ into $ {\bf k_2}$. 
 Simultaneously all 
$z$-components are stretched in the $|k_z/k_{z1}|$ times, 
and all transverse components get the factor $|k_{\bot}/k_{1\bot}|$. 
Besides, the rotation on the angle $\mbox{arg}(\zeta/\zeta_1)$
around $z$-axis takes place.

As the result, due to the properties of both $U$ and $T$, 
$$
U_{k_1|kk_2} \to |k_z/k_{z1}|^{+1}|k_{\bot}/k_{1\bot}|^{-2}U_{k|k_1k_2},\,\,\,
T_{k_1|kk_2} \to |k_z/k_{z1}|^{2\alpha}|k_{\bot}/k_{1\bot}|^{2\beta}T_{k| k_1k_2},
$$
the integrand in (\ref{23}) is factorized:
\begin{displaymath}
\int U_{k|k_1k_2}T_{k|k_1k_2}
\Biggl [  1 - \Bigl (  \frac{k_z}{k_{z1}} \Bigr )^{2\alpha +4}
\Bigl (  \frac{k_{\bot}}{k_{\bot
1}} \Bigr )^{2\beta +4}
\Biggr ] dk_1dk_2 = 0.
\end{displaymath}
Hence it follows that, besides the thermodynamic equilibrium 
spectra (which annulates $T$),  the following solution is possible:
\begin{equation}
\label{25}
n_k=Ak^{-2}_zk^{-2}_{\bot}, \qquad
N_k=Bk^{-2}_zk^{-2}_{\bot}.
\end{equation}
These spectra correspond to the solution which was obtained previously 
from the dimensional analysis for the constant flux of HF waves  $P_N$.
Connection between the coefficients 
 $A$ and $B$ in (\ref{25}) is determined from solution of the stationary
 ($\partial/\partial t=0$) equation  (\ref{18}). 
Hence one can get that for this case total 
energies containing in Alfvenic waves
and in slow magneto-acoustic waves are of the 
same order of magnitude: $c_sA\sim V_AB$. 

It is worth to note that the whole set of 
the transformations (\ref{24}) forms the group  $G$.
This group is direct product of two groups  $G(1)$ and 
$G(2)$: $G=G(1)\times G(2)$. The group  $G(1)$ acts in one-dimensional space
($k_z$-space), and $G(2)$  in the two-dimensional one ($k_{\bot}$-space). 
 These transformations allow one to factorize the collision terms.
 First these transformations (in the 1D case of the frequency space) 
 were found by V.E.Zakharov \cite{Z2,ZF,Z3}. 
The generalizations of these transformation to the both 2D and 3D cases for isotropic 
 models were introduced  by A.V.Kats and V.M.Kontorovich 
 in 1970 \cite{kk}. The transformations (\ref{24}) were found by the author in 1972
 \cite{kuz}. They represent a partial type of the so-called quasi-conformal
 transformations.

 To find another non-equilibrium solution of (\ref{19'}) 
it is convenient to introduce the energy 
density in the $k$-space
$\varepsilon_k = \omega_kN_k+\Omega_kn_k$.
From (\ref{21}) and  (\ref{21'}) follows that this quantity obeys 
the equation:
\begin{equation}
\label{26}
\frac{\partial\varepsilon_k}{\partial t} =
\int \Bigl \{ \omega_k U_{k|k_1k_2}T_{k|k_1k_2}-
\omega_k U_{k_1|kk_2}T_{k_1|kk_2} 
-\Omega_kU_{k_2|k_1k}T_{k_2|k_1k} \Bigr \} dk_1dk_2\,.
\end{equation}
Consider stationary solution of this equation. As before, we shall seek
for solution of (\ref{26}) in the form (\ref{22}). In this case we have three
integrals, integration areas of which are defined by the 
appropriate $\delta$-functions.
Therefore we shall make transformations analogous to (\ref{24}).
The transformations of the integration area of the second 
integral standing in (\ref{26})
will be the same as (\ref{24}). The transformation of 
the integration area of the third 
integral into the corresponding integration area of the 
first integral in (\ref{26})
has the form:
\begin{equation}
\label{27}
k_z = k_z'' \frac{k_z}{k_z''}, \qquad  \zeta=\zeta'' \frac{\zeta}{\zeta''},
\end{equation}
\begin{displaymath}
k_{z_1} = k_z' \frac{k_z}{k_z''},  \qquad  \zeta_1=\zeta' 
\frac{\zeta}{\zeta''},
\end{displaymath}
\begin{displaymath}
k_{z_2} = k_z \frac{k_z}{k_z''},  \qquad  \zeta_2=\zeta 
\frac{\zeta}{\zeta''}.
\end{displaymath}

Applying all of these transforms yields factorization of the integrand 
for the stationary equation (\ref{26}):
\begin{displaymath}
0=\int | V_{kk_1k_2} |^2
\delta_{k-k_1-k_2}\delta_{\omega -\omega_1-\Omega_2}
T_{k|k_1k_2}dk_1dk_2\cdot
\end{displaymath}
\begin{displaymath}
\Biggl\{ \omega(k)-\omega(k_1)
\Bigl ( \frac{k_z}{k_{z_1}} \Bigr )^{2\alpha +5}
\Bigl ( \frac{k_{\bot}}{k_{\bot_1}} \Bigr )^{2\beta +4}-
\Omega(k_2) \Bigl ( \frac{k_z}{k_{z_2}} \Bigr )^{2\alpha +5}
\Bigl ( \frac{k_{\bot}}{k_{\bot_2}} \Bigr )^{2\beta +4}
\Biggr \}.
\end{displaymath}
Hence it follows that the figure bracket vanishes at
$\alpha = -5/2,~ \beta = -2$, namely, the solution has the form 
\begin{equation}
\label{energy}
n_k = Ak^{-5/2}_zk^{-2}_{\bot}, \qquad
N_k = Bk^{-5/2}_zk^{-2}_{\bot} \,.
\end{equation}
The obtained solution corresponds to spectra with constant energy flux 
 $P_{\varepsilon}$. Connection between constants 
$A$ and $B$, as before,  is found from the stationary equation  (\ref{18}).
From this equation it is possible to get the previous estimation on their ratio:
$c_sA\sim V_AB$.

The found above solutions of the Kolmogorov type are related to the 
only channel of interaction, namely, the interaction between Alfvenic and 
slow magneto-acoustic waves that demanishes significantly the value of these solutions. Remind, 
that processes together with fast magneto-acoustic waves have  growth rates of 
the same order of magnitudes, and therefore they can not be ignored. 
Fortunately, 
the channel (with the fast magneto-acoustic waves) 
can be incorporated
in the considered above scheme without essential generalizations.
 As was pointed out in the previous
section, the maximal scattering of A-waves is attained at 
the maximal value of $z$-projection of
momentum transfered to slow magneto-acoustic waves while scattering 
of A-waves. Therefore it is natural to 
assume that such behavior of the scattering amplitude of A-waves  
should lead to 
strongly anisotropic distribution of waves concentrated  in a narrow 
cone of angles along mean magnetic field: $k_z\gg k_{\bot}$. 
Under this assumption it is possible to consider 
the dispersion law of fast magneto-acoustic
waves to be approximated by those for Alfvenic waves: 
$\omega_2\approx |k_z|V_A$. Another important circumstance 
in this case is that 
the matrix element of interaction is diagonal with respect to polarization
 $\lambda$:
$$
V^{\lambda \lambda_1}_{k'k'_1k_2}
\approx\delta_{\lambda \lambda_1}V^{11}_{k'k'_1k_2}.
$$
Thus, for almost longitudinal (along the mean magnetic field) 
distribution 
there is almost no difference between Alfvenic and fast 
magneto-acoustic waves. Moreover, 
the direct energy exchange is absent between these waves. 
This means that for 
this region of angles {\it Kolmogorov spectra for fast magneto-acoustic waves}
will have the same form as those of the obtained spectra 
 (\ref{25}) and (\ref{energy}). In this case in the expressions 
(\ref{25}) and (\ref{energy})
$N_k$ and $B$ should be changed into $N_k^{\lambda}$ and $B_{\lambda}$, 
and the coefficient
$$
A \sim \beta^{-1/2}\frac{\sum B_{\lambda}^2}{\sum B_{\lambda}}.
$$
  
The spectra, obtained in this section, will have the physical meaning 
if the locality property will be
fulfilled. This requirement of locality consists in that 
contributions into interaction  of the waves
from both the pumping region and the dissipation region have 
to be small. The latter leads to the
requirement of convergence of integrals in the equations 
(\ref{21}) and (\ref{21'}).

Convergence of integrals relative to $k_z$ provides by the presence of two 
$\delta$-functions containing $k_z$. As far as convergence against
transverse wave vectors concerns, the integrals are logarithmically divergent.
The logarithmical divergence, to our opinion, is not so serious as a possible 
powerful one. Appearance of divergence is connected with bi-homogeneity of 
the probability $U$.  If a medium would be isotropic and matrix elements 
 $V$ would would have the same degrees of homogeneity as for the MHD waves at  $\beta\ll 1$ 
(such situation, for instance, takes place for Mandelstamm-Brillouin 
scattering in 
isotropic dielectrics) then in such a case the locality property would be valid
(compare with \cite{ZK78}). Violation of bi-homogeneity for 
interaction of the Alfvenic and slow
magneto-acoustic waves appears for almost transverse propagation:
 $k_{\bot}/k_z\sim \beta^{-1/2}$ and for interaction of the fast 
magneto-acoustic waves
 for small angles $\sim \beta^{1/2}$. By this reason cut-off of integrals in 
 the kinetic equations should be performed on the smaller angles: 
$\sim \beta^{1/2}$.
 Another possibility to avoid the logarithmic divergence 
 is in seeking for solutions containing powers of logarithm from
 $k_{\bot}$  in (\ref{25})  and (\ref{energy}). 
However, this procedure does not lead to determination of the powers, 
but, however, provides a 
convergence of the integrals. 

Last remark. The spectra (\ref{25}) as well as (\ref{energy}) 
have the same power dependence on transverse momenta:
$n_k,N_k\sim k_{\bot}^{-2}$. Their homogeneity degree against transverse momenta
is the same as for 2D
$\delta$-function of ${\bf k_{\bot}}$. This means that, besides the anisotropic 
Kolmogorov spectra  (\ref{25}) and (\ref{energy}), the
singular Kolmogorov spectra are possible:
$$
n_k = Ak^{-2}_z\delta({\bf k_{\bot}}), \qquad
N_k = Bk^{-2}_z\delta({\bf k_{\bot}})
$$
and 
$$
n_k = Ak^{-5/2}_z\delta({\bf k_{\bot}}), \qquad
N_k = Bk^{-5/2}_z\delta({\bf k_{\bot}}).
$$
Which spectra are realized indeed? Rigorous answer to this question is possible 
to get by stability investigation of the spectra or 
numerical experiment (in the latter case one has a hope on the qualitative understanding).  
Both these approaches require separate consideration.

\section*{Acknowledgments}
The author thanks to V.E.Zakharov for many fruitful discussions.
The author is thankful also to the Nice Observatory, where 
this paper was completed within 
Landau-CNRS agreement, for their hospitality.
This work was supported by the RFBR under Grant No. 00-01-00929),
by the Program of Support of the Leading Scientific Schools of Russia 
(Grant No. 00-15-96007) and by the INTAS (Grant No. 00-00292).

\end{document}